\documentclass[reprint,aps,prl,amsmath,amssymb,twocolumn,notitlepage,longbibliography,floatfix]{revtex4-1}
\usepackage{amsmath,amssymb, bm,graphicx}
\usepackage{graphics}
\usepackage[stable]{footmisc}
\usepackage[colorlinks, linkcolor= blue, citecolor = blue, urlcolor=blue]{hyperref}
\usepackage{xcolor}
\usepackage{physics}
\usepackage{chemformula}
\usepackage{soul}
\usepackage{physics}
\usepackage{pifont}
\usepackage{gensymb}
\usepackage{tabularx}    
\usepackage{pifont}      
\usepackage{array} 
\usepackage{orcidlink}
\newcommand{\cmark}{\ding{51}}
\newcommand{\xmark}{\ding{55}}

\def\be{\begin{equation}}
\def\ee{\end{equation}}
\def \bea{\begin{eqnarray}}
\def \eea{\end{eqnarray}}
\def \nn{\nonumber}

\begin{document}
\title{Intrinsic Magnetoelectric Hall Effect from Layer-Orbital Quantum Geometry}

\author{Sunit Das}
\email{sunitd@iitk.ac.in}
\author{Amit Agarwal}
\email{amitag@iitk.ac.in}
\affiliation{Department of Physics, Indian Institute of Technology Kanpur, Kanpur 208016, India}

\begin{abstract}
Intrinsic Hall effects, such as the anomalous Hall effect, originate from the orbital quantum geometry of Bloch states. However, in layered materials, the combined action of out-of-plane electric and magnetic fields couples to layer polarization and orbital moment, generating a mixed layer-orbital quantum geometry in field-dressed Bloch states. We show that this geometry produces an intrinsic magnetoelectric Hall effect that is bilinear in the electric and magnetic fields. The response is scattering-time independent and can arise in nonmagnetic systems without spin-orbit coupling. Its origin lies in interband coherence involving layer polarization and orbital moment, leading to a finite, non-quantized Hall response that persists in the band gap. The Hall coefficient is odd under gate reversal and tracks layer polarization. A symmetry analysis identifies the classes of layered materials that host this effect. As a representative realization, we demonstrate the effect in rhombohedral pentalayer graphene, where the conductivity reaches values of order $0.05\,e^2/h$. These results establish mixed layer-orbital quantum geometry as a mechanism for intrinsic magnetoelectric Hall transport and a direct probe of layer-resolved quantum geometry in Bloch bands.
\end{abstract}

\maketitle
\clearpage
\textit{Introduction---}Intrinsic Hall responses are a central manifestation of the quantum geometry of Bloch bands in solids~\cite{xiao_rmp10_berry, Su-Yan_25, Verma2026, Yang_25_QG, Xie_nsr, Adak2024, Dimi}. Beyond the Berry curvature-driven anomalous Hall effect, this geometric viewpoint now underlies a broader class of nonlinear and field-induced transport phenomena governed by the quantum metric and related quantities~\cite{Carmine_aqm21, BANDYOPADHYAY2024100101, Binghai25, Heuso}. These developments establish band geometry as a unifying framework for intrinsic Hall transport. However, they primarily focus on the band geometry associated with the orbital degrees of freedom. The central open question is whether external fields can generate new intrinsic Hall responses by inducing geometric structure in additional internal degrees of freedom of Bloch states~\cite{Wang_prl25}.

Layered van der Waals materials provide a natural platform for addressing this question, as their Bloch states host a layer degree of freedom that can be tuned by an out-of-plane electric field. This layer degree of freedom already underlies a range of layer-resolved transport phenomena~\cite{Gao2021, Han_prl25, Feng2023, Chen_2025, Yandong}. At the same time, transport responses driven separately by electric and magnetic fields have been extensively studied in layered systems~\cite{Ghorai_prl25, Wang_NL24, Wang_NC23, Wang_NL25, Wang_prrL24, Long_Nat23, Justin_25, hu2025, Yao_prl24_electric, Wang_prb24, Xie_nsr23, Wang_prrL24}. What remains unresolved is whether their combined action can induce a new intrinsic Hall response. In particular, can field-driven mixing of layer polarization and orbital moment generate an intrinsic Hall effect without relying on spin–orbit coupling or magnetic order?

In this Letter, we show that simultaneous out-of-plane electric and magnetic fields generate a mixed layer-orbital quantum geometry in field-dressed Bloch states. Electric fields polarize layers, while magnetic fields couple to orbital motion, and their interplay gives rise to interband coherence that mixes layer polarization and orbital motion. This produces an intrinsic magnetoelectric Hall effect (IMHE) in a standard Hall-bar geometry. The response is scattering-time independent, bilinear in the out-of-plane fields ${\cal E}$ and $B$, and admits an equivalent description in terms of an ${\cal E}B$-induced orbital magnetization. We determine the symmetry conditions under which it is allowed and demonstrate it in rhombohedral pentalayer graphene. These results establish mixed layer-orbital quantum geometry as a new mechanism for intrinsic Hall transport in layered systems.
\begin{figure}[t]
    \centering
    \includegraphics[width=\linewidth]{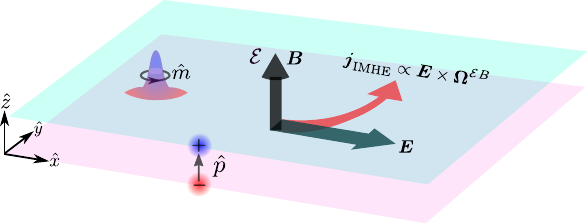}
    \caption{\textbf{Intrinsic magnetoelectric Hall effect from layer–orbital quantum geometry.} 
    In multilayer systems, an out-of-plane electric field ${\cal E}\hat{z}$ polarizes layers and couples to the layer dipole moment $\hat{p}$, while a magnetic field $B\hat{z}$ couples to the orbital magnetic moment $\hat m$. Their combined action generates interband coherence that mixes layer polarization and orbital motion, producing a field-induced layer–orbital contribution to the Berry curvature, $\boldsymbol{\Omega}^{{\cal E}B}$. This mixed geometric term gives rise to an intrinsic Hall response $\propto {\cal E}B$ in a standard Hall-bar geometry.
    \label{fig1}}
\end{figure}

\textit{Layer-orbital quantum geometry---}To describe an intrinsic Hall response under simultaneous out-of-plane electric and magnetic fields, we track how these fields reshape the band geometry of the Bloch states. In a layered system, the gate electric field ${\cal E}\hat{z}$ couples to the layer polarization operator $\hat{p}$, while the magnetic field $B\hat{z}$ couples to the orbital moment operator $\hat{m}$ (see Fig.~\ref{fig1}). The Hamiltonian is then given by
\be \label{Ham_layer}
\hat{\cal H} = \hat{\cal H}_0 + \hat{p}\,{\cal E} - \hat{m}\,B~,
\ee
where $\hat{\cal H}_0\ket{u_{n\bm k}}=\varepsilon_{n\bm k}\ket{u_{n\bm k}}$, with $\ket{u_{n\bm k}}$ and $\varepsilon_{n\bm k}$ denoting the unperturbed Bloch eigenstate and eigenenergy of the $n$-th band. 

The combined action of the out-of-plane ${\cal E}$ and $B$ generates interband coherence that mixes layer and orbital character in the field-dressed Bloch states. Treating ${\cal E}$ and $B$ as weak perturbations, the field-modified Bloch state is
\be \label{field_state}
|\tilde{u}_{n\bm k}\rangle = |u_{n\bm k}\rangle
+ {\cal E} \sum_{q\neq n} \frac{p_{qn}}{\varepsilon_{nq}} |u_{q\bm k}\rangle
- B \sum_{q\neq n} \frac{m_{qn}}{\varepsilon_{nq}} |u_{q\bm k}\rangle~.
\ee
Here, $\varepsilon_{nq} = \varepsilon_{n\bm k} - \varepsilon_{q\bm k}$, and
$p_{qn}=\langle u_{q\bm k}|\hat{p}|u_{n\bm k}\rangle$ and
$m_{qn}=\langle u_{q\bm k}|\hat{m}|u_{n\bm k}\rangle$ are the corresponding interband matrix elements.

This field-induced interband mixing reshapes the band geometry. In particular, the Berry curvature acquires corrections,
\be \label{BC_corr}
\tilde{\bm \Omega}_{n\bm k}
= {\bm \Omega}_{n\bm k}
+ {\cal E}\,{\bm \Omega}^{\cal E}_{n\bm k}
+ B\, {\bm \Omega}^{B}_{n\bm k}
+ {\cal E}B\, {\bm \Omega}^{{\cal E}B}_{n\bm k}~.
\ee
The mixed term ${\bm \Omega}^{{\cal E}B}_{n\bm k}$ is the key band geometric quantity generated by the coupled fields. Microscopically, it is governed by products of interband matrix elements of $\hat{p}$ and $\hat{m}$ weighted by band-energy denominators, with representative contributions such as $p_{nq} m_{qn}/\varepsilon_{nq}$ (see Appendix~A). Thus, we identify it as a {\it layer-orbital polarizability} to Berry curvature, a field-induced geometric response arising from interband coherence between the layer and orbital sectors. It differs from the purely gate electric correction ${\bm \Omega}^{\cal E}_{n\bm k}$~\cite{Yao_prl24_electric} and from the purely magnetic correction ${\bm \Omega}^{B}_{n\bm k}$~\cite{Yang_prl24}, and it appears only under their combined action. An analogous mixed correction also appears in the orbital magnetic moment $\tilde{\bm m}_{n\bm k}$.

The band energy is corrected analogously,
\begin{equation}
\tilde{\varepsilon}_{n\bm k}
= \varepsilon_{n\bm k}
+ {\cal E}\,\varepsilon^{\cal E}_{n\bm k}
+ B\,\varepsilon^{B}_{n\bm k}
+ {\cal E}B\,\varepsilon^{{\cal E}B}_{n\bm k}~.
\end{equation}
Here, $\varepsilon^{{\cal E}}_{n\bm k}= -e d p_{n\bm k}$ and $\varepsilon^{B}_{n\bm k}=-m_{n\bm k}$ are the Stark-like and orbital Zeeman contributions, respectively. $p_{n\bm k}$ is the intraband layer dipole moment, and $d$ denotes the effective vertical separation associated with layer polarization, i.e., the top-to-bottom layer spacing. $\varepsilon^{{\cal E}B}_{n\bm k}$ is associated with the corresponding magnetoelectric coupling energy~\cite{Justin_25}. Explicit expressions for ${\bm \Omega}^{{\cal E}B}_{n\bm k}$, ${\bm m}^{{\cal E}B}_{n\bm k}$, and $\varepsilon^{{\cal E}B}_{n\bm k}$ are given in Appendix~A.

These field-induced geometric quantities are gauge-invariant, modify carrier dynamics, and give rise to observable transport phenomena. In particular, ${\bm \Omega}^{{\cal E}B}_{n\bm k}$ generates an intrinsic magnetoelectric Hall response under an in-plane bias (see Fig.~\ref{fig1}) that is bilinear in the out-of-plane fields and remains finite in nonmagnetic layered systems. 

\textit{Intrinsic magnetoelectric Hall effect---}We now derive the intrinsic Hall response generated by this field-induced geometry. Within semiclassical Boltzmann theory, the charge current is ${\bm j} = -e \int_{n\bm k}  \dot{\bm r}_{n\bm k}\, g_{n\bm k}$,
where $g_{n\bm k}$ is the nonequilibrium distribution function, $\dot{\bm r}_{n\bm k}$ is the wave-packet velocity, and $\int_{n\bm k}\equiv \sum_n \int [d{\bm k}]$ with $[d{\bm k}] \equiv d^2{\bm k}/(2\pi)^2$ for a two-dimensional system. Including the ${\cal E}B$-induced corrections in the Berry curvature, orbital magnetic moment, band velocity, and distribution function, we obtain the current linear in the in-plane bias and bilinear in the out-of-plane fields. The magnetoelectric current proportional to ${\cal E}B$ is
\bea \label{current_generic}
j_a = \sigma_{ab} E_b \equiv \chi_{ab;zz}^{\rm in} E_b {\cal E} B + \tau \chi_{ab;zz}^{\rm ex} E_b {\cal E} B~.
\eea 
Here, $E_b$ is the in-plane bias field, $\sigma_{ab}$ is the effective magnetoelectric conductivity, $\tau$ is the scattering time, and $\{a,b\}\in\{x,y\}$ label the in-plane directions. The first term represents the IMHE and is our main focus. The second term is an extrinsic correction proportional to $\tau$, included for completeness. 

The intrinsic response is
\bea \label{chi_hall}
\chi_{ab;zz}^{\rm in} &=& -\frac{e^2   }{\hbar} \epsilon_{abz} \int_{n\bm k} \left[  {\Omega}^{{\cal E} B} \, f_0 +  {\Omega} \,     \varepsilon^{{\cal E} B} \, \partial_{\varepsilon}f_0 \right]~.
\eea 
Here, $\epsilon_{abz}$ is the Levi-Civita tensor and $f_0 \equiv f_0(\varepsilon)$ is the equilibrium Fermi-Dirac distribution function. For brevity, band and momentum labels are suppressed. Equation~\eqref{chi_hall} is the central result of this work. It shows that the intrinsic IMHE is governed entirely by the ${\cal E}B$-induced geometric corrections. The first term arises from the field-induced Berry curvature ${\Omega}^{{\cal E}B}$ and represents a Fermi-sea contribution, while the second term originates from the magnetoelectric energy correction $\varepsilon^{{\cal E}B}$ and captures a Fermi-surface response weighted by the equilibrium Berry curvature.

For completeness, the extrinsic contribution associated with $\varepsilon^{{\cal E}B}$ is
\be \label{chi_lon}
\chi_{ab;zz}^{\rm ex}
= -{e^2}\!\int_{n\bm k}\!
\left[ v_a^{{\cal E}B}\,v_b +(a \leftrightarrow b) \right] \partial_\varepsilon f_0
+ v_a\,v_b\,\varepsilon^{{\cal E}B} \, \partial_\varepsilon^2 f_0~, 
\ee
where the field-induced velocity correction satisfies $\hbar v_a^{{\cal E}B}=\partial_{k_a}\varepsilon^{{\cal E}B}$ and the unperturbed band velocity is $\hbar v_a=\partial_{k_a}\varepsilon$. Additional terms from combining the separate ${\cal E}$- and $B$-induced corrections to the energy and quantum geometry are given in Sec.~S1 of Supplemental Material (SM)~{\footnote{See Supplemental Material at [link to be inserted by the publisher] for details on (S1) derivation of magnetoelectric current expressions, (S2) magnetic point group symmetry restriction, (S3) magnetoelectric transport response in R5G, and (S4) distinction of IMHE from other Hall mechanisms.}}, along with the derivation of Eq.~\eqref{chi_hall} and \eqref{chi_lon}. 

The intrinsic tensor $\chi^{\rm in}_{ab;zz}$ is antisymmetric under $a\leftrightarrow b$. Therefore, it produces a purely transverse Hall current with ${\bm j} \cdot {\bm E}=0$. By contrast, the extrinsic tensor $\chi^{\rm ex}_{ab;zz}$ is symmetric and yields dissipative longitudinal and transverse contributions with ${\bm j} \cdot {\bm E}\neq0$. For an in-plane field along $\hat{x}$, the intrinsic IMHE conductivity is $\sigma^{\rm IMHE}=\chi^{\rm in}_{yx;zz}{\cal E}B$, whereas the extrinsic conductivities are $\sigma^{\parallel(\perp)}=\tau \chi^{\rm ex}_{xx(yx);zz}{\cal E}B$.

The intrinsic IMHE contains a Fermi sea contribution and originates from the field-induced mixing of layer polarization and orbital motion encoded in ${\Omega}^{{\cal E}B}$ and $\varepsilon^{{\cal E}B}$. Unlike anomalous Hall effects, it does not rely on a topological invariant and is therefore non-quantized in the band gap, with its magnitude set by material-specific band geometry. This establishes the IMHE as a distinct intrinsic Hall mechanism arising from mixed layer-orbital quantum geometry, beyond both Berry curvature-driven anomalous Hall responses and conventional Lorentz transport~\cite{Note1}. We show below that this response is tied to ${\cal E}B$-induced nonequilibrium orbital magnetization and the associated boundary currents rather than topological edge states.

\textit{IMHE from nonequilibrium orbital magnetization---}The ${\cal E}B$-induced layer-orbital polarizability provides a geometric description of the IMHE in nonmagnetic layered materials (see Table~\ref{table_1}). An equivalent physical picture emerges in terms of an out-of-plane nonequilibrium orbital magnetization $M_z$, which directly drives the response.

The Hall conductivity can be expressed in terms of the orbital magnetization as $\sigma_{ab} = -e \, \epsilon_{abc} \, (\partial_{\mu} {M}_c)$, where $\mu$ is the chemical potential~\cite{jackson_12, Wen_prr25}. This motivates us to evaluate the orbital magnetization per unit area arising from the coupled ${\cal E}B$ correction. We obtain the Magnetization correction as, 
\bea  \label{Orb_Mag_EB}
{M}_z ({\cal E} B)  = -\frac{e}{\hbar} {\cal E} B \int_{n\bm k} \left[\Omega^{{\cal E}B} \, g(\varepsilon) - \Omega \, \varepsilon^{{\cal E}B} f_0(\varepsilon) \right]~,
\eea 
with $\partial_\mu g(\varepsilon) = -f_0(\varepsilon)$. Using Eq.~\eqref{Orb_Mag_EB}, we recover the IMHE response of Eq.~\eqref{chi_hall}. Physically, the ${\cal E}B$-induced orbital magnetization generates circulating currents and thus counter-propagating boundary currents at the sample edges. In equilibrium ($E=0$), these currents cancel between opposite edges. An in-plane electric field shifts the chemical potential locally, unbalancing the boundary currents and generating a net transverse voltage (see Fig.~S1 of SM~\cite{Note1}). This reveals that the intrinsic IMHE can be interpreted as the transport consequence of an ${\cal E}B$-induced nonequilibrium orbital magnetization. See Appendix~B for details, including the derivation of Eq.~\eqref{Orb_Mag_EB}. As a bulk thermodynamic contribution, the IMHE is not associated with topological edge states. With this physical picture in place, we turn to the symmetry conditions under which the IMHE is allowed.

\begingroup
\setlength{\tabcolsep}{4pt}    
\renewcommand{\arraystretch}{1.1}  
\begin{table}[t]
\centering
\small
\caption{{\bf Symmetry restrictions on the magnetoelectric response tensors.} Crosses (\xmark) and ticks (\cmark) denote symmetry-forbidden and symmetry-allowed responses, respectively. Here, ${\cal M}_{a}$, ${\cal C}_{na}$, and ${\cal S}_{na}$ denote mirror, $n$-fold rotation, and $n$-fold roto-inversion operations about $a$-axis with $a=\{x,y,z\}$.}
\label{table_1}
\begin{tabular}{lcc}
\hline\hline
{Symmetry elements} & $\chi^{\mathrm{in}}_{yx;zz}$  & $\chi^{\mathrm{ex}}_{xx;zz}$  \\
\hline

$\mathcal{P},\ \mathcal{P}\mathcal{T},\ \mathcal{M}_z,\ \mathcal{S}_{6z},\ \mathcal{C}_{2x}\mathcal{T},\ \mathcal{C}_{2y}\mathcal{T}$ 
   & \xmark  & \xmark  \\[6pt]

$\mathcal{T},\ \mathcal{M}_x,\ \mathcal{M}_y,\ \mathcal{C}_{2z}\mathcal{T},\ \mathcal{C}_{3z}\mathcal{T},\ \mathcal{C}_{6z}\mathcal{T}$
   & \cmark  & \xmark \\[6pt]

$\mathcal{C}_{2x},\ \mathcal{C}_{2y},\ \mathcal{S}_{4z},\ \mathcal{M}_z\mathcal{T},\ \mathcal{C}_{2z}\mathcal{T},\ \mathcal{C}_{3z}\mathcal{T},\ \mathcal{C}_{6z}\mathcal{T}$
   & \xmark & \cmark  \\[6pt]

$\mathcal{C}_{2z},\ \mathcal{C}_{3z},\ \mathcal{C}_{4z},\ \mathcal{C}_{6z},\ \mathcal{M}_x\mathcal{T},\ \mathcal{M}_y\mathcal{T}$,  $\mathcal{C}_{4z}\mathcal{T}$
   & \cmark  & \cmark \\

\hline\hline
\end{tabular}
\end{table}
\endgroup

\begin{figure}[t]
    \centering
    \includegraphics[width=\linewidth]{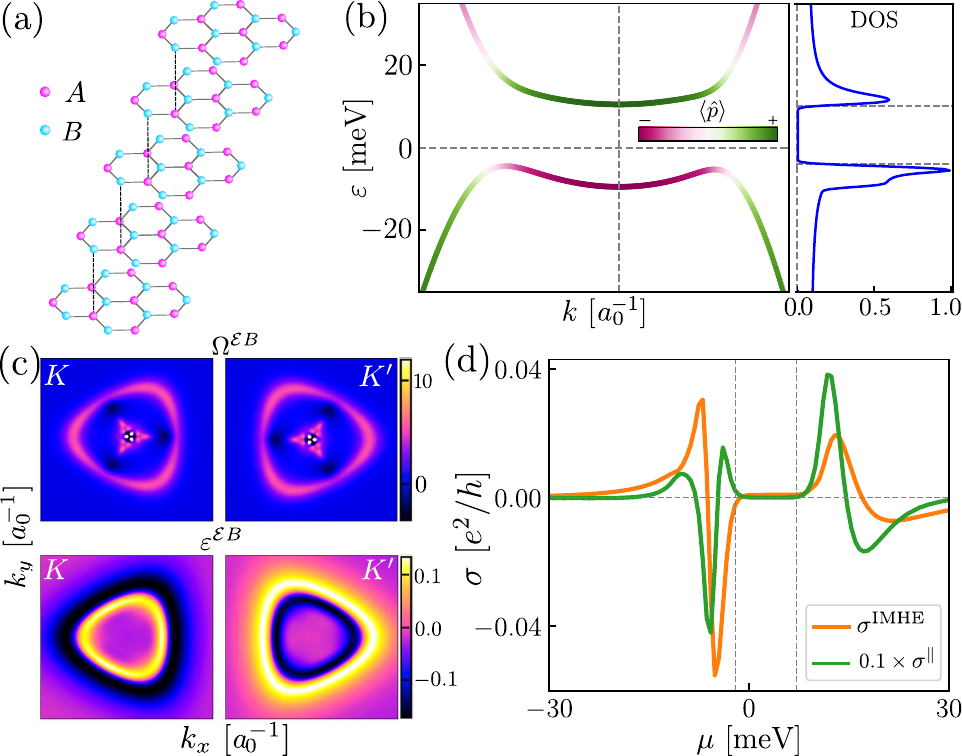}
    \caption{{\bf IMHE response in rhombohedral pentalayer graphene.} 
    (a) Side view of the atomic stacking in rhombohedral pentalayer graphene (R5G), with $A$- and $B$-type atoms indicated. 
    (b) Layer-composition-weighted low-energy band structure at the $K$ valley for a gate potential $\Delta_{\cal E}=20~\rm meV$ and a valley polarization of $0.5~\rm meV$, shown together with the normalized density of states. $a_0$ is the lattice constant. 
    (c) Layer-orbital polarizability of the Berry curvature $\Omega^{{\cal E}B}$ (in units of $ e \,{\rm nm}^5/\hbar$) and energy correction $\varepsilon^{{\cal E}B}$ (in units of $ e^2 \, {\rm nm}^3 /\hbar$) for the first valence band in the two valleys. 
    (d) Intrinsic magnetoelectric Hall conductivity as a function of $\mu$, together with the longitudinal response $\sigma^{\parallel}$ (for $\tau=0.1$ ps) at $T=5$ K. }
    \label{fig2}
\end{figure}

\textit{Crystalline symmetry restrictions---}We now identify the symmetry conditions under which the magnetoelectric response is allowed. Inversion symmetry $\cal P$ must be broken because both $\chi^{\rm in}_{ab;zz}$ and $\chi^{\rm ex}_{ab;zz}$ describe a second-order electric-field response proportional to ${\cal E}E$. The horizontal mirror ${\cal M}_z$ also forbids both tensors: under ${\cal M}_z$, ${\cal E}$ changes sign whereas $j$, $B$, and $E$ remain unchanged [see Eq.~\eqref{current_generic}].

The intrinsic tensor $\chi^{\rm in}_{ab;zz}$ is a $\cal T$-even axial tensor~\cite{newnham_symmetry,Gallego_cryst19}. It can therefore be finite in nonmagnetic materials, but it vanishes in $\cal PT$-symmetric systems such as bipartite antiferromagnets. By contrast, the extrinsic tensor $\chi^{\rm ex}_{ab;zz}$ is $\cal T$-odd and therefore requires broken time-reversal symmetry. Table~\ref{table_1} summarizes the symmetry-allowed components of $\chi^{\rm in}_{yx;zz}$ and $\chi^{\rm ex}_{xx;zz}$ for representative point-group elements; a complete classification of all 122 magnetic point groups, including $\chi^{\rm ex}_{yx;zz}$, is provided in Sec.~S2 of SM~\cite{Note1}. These constraints identify nonmagnetic layered systems with broken inversion and horizontal mirror symmetry as a natural materials platform for the IMHE.

Notably, ${\cal M}_x$, ${\cal M}_y$, and the threefold rotation ${\cal C}_{3z}$ do not forbid the intrinsic IMHE. These symmetry conditions arise naturally in hexagonal layered systems such as multilayer graphene and moir\'e heterostructures. The extrinsic longitudinal response $\chi^{\rm ex}_{xx;zz}$ is allowed only when all mirror symmetries are absent. We now illustrate the IMHE response in rhombohedral pentalayer graphene as a concrete realization.

\textit{IMHE in rhombohedral pentalayer graphene---}Rhombohedral pentalayer graphene (R5G) is a natural platform for the IMHE. Its low-energy states are strongly polarized on the two outermost layers, so gate fields act efficiently on the layer degree of freedom. The ABC-stacked structure [Fig.~\ref{fig2}(a)] lacks inversion symmetry~\cite{Koshino_prb09, Knolle_25, Parker_prl24, Senthil_prl24, Fu_prl25}, while a gate field or substrate breaks the ${\cal M}_z$ mirror symmetry, satisfying the symmetry conditions derived above. Under these conditions, the intrinsic Hall tensor $\chi^{\rm in}_{yx;zz}$ is allowed. Recent observations of spontaneous valley polarization and orbital magnetism in R5G~\cite{Long_Nat23, Long_sc24} further indicate that time-reversal symmetry can be broken, enabling extrinsic responses. Yet ${\cal C}_{3z}$ forbids the transverse extrinsic response $\chi^{\rm ex}_{yx;zz}$ (see Sec.~S2 of SM~\cite{Note1}), while the intrinsic Hall component $\chi^{\rm in}_{yx;zz}$ is allowed. R5G therefore isolates $\chi^{\rm in}_{yx;zz}$ as the only magnetoelectric transverse response.

To model the band structure of R5G, we use a $10\times 10$ low-energy continuum Hamiltonian~\cite{Koshino_prb09, Senthil_prl24}, as detailed in Sec.~S3 of SM~\cite{Note1}. Figure~\ref{fig2}(b) shows the low-energy bands near the $K$ valley for a gate field-induced interlayer potential $\Delta_{\cal E}\approx e{\cal E}d/\epsilon_r = 20$ meV, where $d\approx 1.4$ nm denotes the top-to-bottom layer separation of R5G and $\epsilon_r \approx 2.5$ is an effective dielectric constant~\footnote{In multilayer graphene, the interlayer potential drop is renormalized by electrostatic screening arising from layer charge redistribution. As a result, the effective potential difference can become a nonlinear function of the applied gate field and depends on carrier density~\cite{Justin_25}. For simplicity, we neglect these screening effects and treat $\epsilon_r$ phenomenologically.}. We include a constant valley polarization of $0.5$~meV~\cite{Note1}. The bands are weighted by their layer character, and those near charge neutrality are strongly concentrated on the outermost layers. This layer polarization amplifies the layer-orbital response that drives the IMHE. The relatively flat dispersion also produces a pronounced van Hove peak in the density of states, as seen in Fig.~\ref{fig2}(b). 

\begin{figure}[t]
    \centering
    \includegraphics[width=\linewidth]{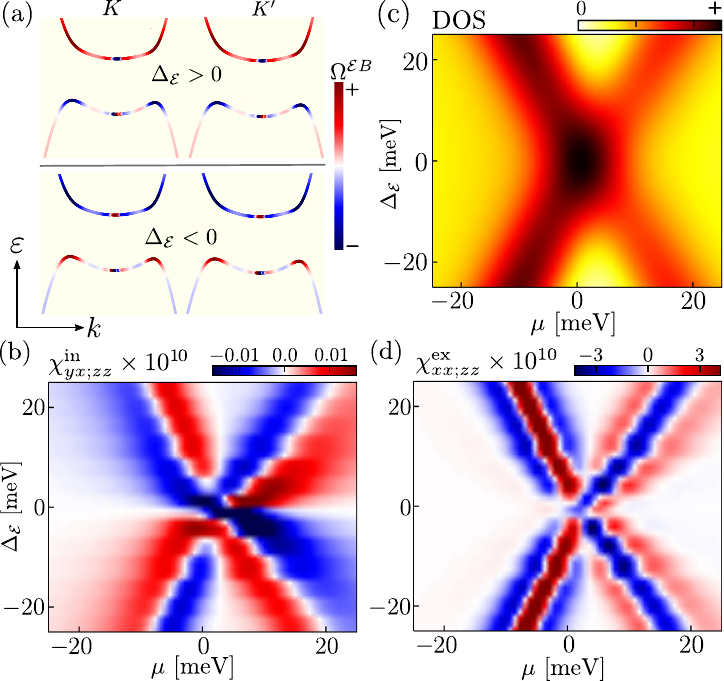}
    \caption{{\bf Gate-voltage tunability of the magnetoelectric response.} (a) $\Omega^{{\cal E}B}$-weighted low-energy band structure of R5G. Reversing the gate field flips the sign of $\Omega^{{\cal E}B}$ through reversal of the layer dipole moment. (b) Intrinsic Hall coefficient $\chi_{yx;zz}^{\rm in}$ (in unit of $\rm AV^{-2}m/T$) as a function of gate-induced potential $\Delta_{\cal E}$ and chemical potential $\mu$ at $T=15$~K. Its sign reversal under $\Delta_{\cal E}\to-\Delta_{\cal E}$ tracks the reversal of the layer polarization. (c) Normalized density of states as a function of $\Delta_{\cal E}$ and $\mu$. (d) Extrinsic longitudinal coefficient $\chi_{xx;zz}^{\rm ex}$ (in unit of $\rm A s^{-1} V^{-2} m/T$) as a function of $\Delta_{\cal E}$ and $\mu$. As a Fermi-surface contribution, it follows the density of states in panel (c) and is strongest near the band edges.}
    \label{fig3}
\end{figure}

The corresponding field-induced band geometry is shown in Fig.~\ref{fig2}(c). The layer-orbital polarizability to Berry curvature $\Omega^{{\cal E}B}$ and the energy shift $\varepsilon^{{\cal E}B}$ are sharply concentrated near the band edges, where the layer polarization is strongest. As required by symmetry, $\Omega^{{\cal E} B}$ is predominantly of the same sign in both valleys [Fig~\ref{fig3}(a)], while $\varepsilon^{{\cal E}B}$ changes sign in two valleys for a fixed $\Delta_{\cal E}$~\footnote{`Predominantly' because the small valley polarization used in the R5G model weakly breaks time-reversal symmetry. The $\Omega^{{\cal E}B}$ ($\varepsilon^{{\cal E}B}$) is $\cal T$-even ($\cal T$-odd). In the absence of valley polarization, consequently, they have the same (opposite) sign and amplitude in the two valleys. See also Sec.~S2 of SM~\cite{Note1}.}. This ties the layer polarization of the bands directly to the geometric quantities that govern the magnetoelectric responses [Eqs.~\eqref{chi_hall} and \eqref{chi_lon}].

We first evaluate the intrinsic response $\chi^{\rm in}_{yx;zz}$. Figure~\ref{fig2}(d) shows the intrinsic Hall conductivity $\sigma^{\rm IMHE}=\chi^{\rm in}_{yx;zz}{\cal E}B$ for ${\cal E} \approx 3.5\times10^6$ V/m and $B=0.1$ T as a function of chemical potential $\mu$. In the band gap, a finite non-quantized response of order $10^{-4}\, e^2/h$ survives and varies with gate voltage [Fig.~\ref{Fig4}(b)]. This behavior is consistent with the nonequilibrium orbital magnetization picture developed in Appendix~B. Across the conduction and valence bands, $\sigma^{\rm IMHE}$ changes sign and reaches peak values of order $0.05\,e^2/h$, corresponding to Hall voltages $\sim \rm mV$ for realistic bias fields~\cite{Note1}. For reference, Fig.~\ref{fig2}(d) also shows the longitudinal response $\sigma^{\parallel}=\chi^{\rm ex}_{xx;zz}{\cal E}B$, which vanishes in the gap and is strongest near the band edges, consistent with its Fermi-surface character.

We next examine the gate-field dependence of the intrinsic response. Figure~\ref{fig3}(a) shows that reversing $\Delta_{\cal E}$ reverses the sign of $\Omega^{{\cal E}B}$ through reversal of the layer dipole moment. Consistently, Fig.~\ref{fig3}(b) shows that the intrinsic IMHE coefficient $\chi^{\rm in}_{yx;zz}$ changes sign under $\Delta_{\cal E}\to-\Delta_{\cal E}$ for all $\mu$. The Hall conductivity $\sigma^{\rm IMHE}=\chi^{\rm in}_{yx;zz}{\cal E}B$, however, remains invariant under gate reversal because the sign change of $\chi^{\rm in}_{yx;zz}$ is compensated by that of ${\cal E}$. The ${\cal E}$-odd coefficient therefore provides a direct transport probe of the underlying layer quantum geometry. For comparison, Fig.~\ref{fig3}(c) and (d) show the density of states and the longitudinal coefficient $\chi^{\rm ex}_{xx;zz}$ as functions of $\Delta_{\cal E}$ and $\mu$. Unlike the intrinsic coefficient, $\chi^{\rm ex}_{xx;zz}$ does not reverse sign under gate inversion and instead follows the field-induced energy correction $\varepsilon^{{\cal E}B}$ together with the associated velocity contributions, which do not change sign under gate-field reversal (see Fig.~S3 of SM~\cite{Note1}).

\textit{Experimental implications---}We now outline how the IMHE can be identified experimentally. Its key signature is set by the measurement geometry and by its parity under field reversal. Conventional intrinsic magneto-Hall responses in layered systems, such as the in-plane anomalous Hall effects~\cite{Zyuzin_prb20, Culcer_prl21, Carmine_prrL21, Liang_NP18, Zhou_nature22, Yang_prl24} or planar Hall effects~\cite{Ghorai_prl25, Wang_NL25}, are induced by in-plane magnetic fields and measured within the plane of the applied fields~\cite{Note1}. Conversely, the IMHE appears in a standard Hall-bar geometry with out-of-plane gate electric and magnetic fields [Fig.~\ref{fig1}], producing a transverse voltage proportional to ${\cal E}B$.

The most direct protocol is a low-frequency lock-in measurement with an AC in-plane bias while sweeping $B$ and ${\cal E}$ independently. Antisymmetrizing the measured signal with respect to $B$ and ${\cal E}$ extracts the bilinear response coefficients $\chi_{yx;zz}^{\rm in}$. The IMHE appears as a transverse voltage odd in both $B$ and ${\cal E}$, whereas the conventional Lorentz Hall signal is even in the gate field (Fig.~S5 of SM~\cite{Note1}). Furthermore, the IMHE exhibits a scattering-time-independent scaling, in contrast to the Lorentz Hall response, which scales distinctly with the longitudinal Drude conductivity (see Sec.~S4 of SM~\cite{Note1}). This distinct scaling, together with the parity under gate-field reversal and the contrasting tensor character of transverse and longitudinal responses, provides a clear experimental fingerprint of the IMHE.

\textit{Conclusion---}We have shown that simultaneous out-of-plane electric and magnetic fields generate a mixed layer-orbital quantum geometry that produces an intrinsic magnetoelectric Hall effect in layered systems. This response is scattering-time independent and arises in nonmagnetic materials without spin–orbit coupling or magnetic order. Its gate-field-odd character provides a direct transport probe of the underlying layer-resolved quantum geometry.

Our symmetry analysis identifies a broad class of layered materials that can host this effect. As a concrete realization, rhombohedral pentalayer graphene exhibits an experimentally accessible conductivity of order $0.05\,e^2/h$. More broadly, these results establish mixed layer-orbital quantum geometry as a general mechanism for intrinsic magnetoelectric transport and provide a pathway for probing geometric responses in layered and moir\'e materials.

\textit{Acknowledgments---}SD acknowledges funding support from the Prime Minister's Research Fellowship under the Ministry of Education, Government of India. AA acknowledges funding from the Core Research Grant by ANRF (Sanction No. CRG/2023/007003), Department of Science and Technology, India.

\bibliography{refs}

\onecolumngrid
\begin{center}
\textbf{\large End Matter}
\end{center}
\twocolumngrid
\renewcommand{\theequation}{A\arabic{equation}}
\renewcommand{\thesection}{A\arabic{section}}
\renewcommand{\thetable}{A\arabic{table}}
\renewcommand{\theHequation}{A\arabic{equation}}
\renewcommand{\theHsection}{A\arabic{section}}
\renewcommand{\theHfigure}{A\arabic{figure}}
\renewcommand{\theHtable}{A\arabic{table}}
\setcounter{table}{0}
\setcounter{equation}{0}
\setcounter{section}{0}

\textit{Appendix A: Field-induced quantum geometry and energy correction---}We consider a general Bloch Hamiltonian $\hat{\cal H}_0$ for a layered material,
$ \hat{\cal H}_0|u_{n{\bm k}} \rangle =  \varepsilon_n({\bm k}) |u_{n {\bm k}}\rangle $, where $|u_{n{\bm k}} \rangle$ is the Bloch eigenstate with energy $\varepsilon_n({\bm k})$. The Berry curvature and orbital magnetic moment of the $n$th band are~\cite{Xiao_prl05}
\be \label{gen_BC_OMM}
{\bm \Omega}_{n\bm k} = i \sum_{m\neq n} {\bm r}_{nm} \times {\bm r}_{mn}~,~~
{\bm m}_{n\bm k} = \frac{e}{2} \sum_{m\neq n} {\bm v}_{nm} \times {\bm r}_{mn}~.
\ee
Here, $\hbar {\bm v}_{nm} = \langle u_{n{\bm k}}|\partial_{k_a} \hat{\cal H}_0|u_{m {\bm k}}\rangle$ denotes the interband velocity matrix element, and ${\bm r}_{nm} = - i \hbar {\bm v}_{nm} / (\varepsilon_n - \varepsilon_m)$ is the interband Berry connection.  

In the presence of the gate electric field $\cal E$ and the magnetic field $B$, the Hamiltonian is given by Eq.~\eqref{Ham_layer}. Treating these fields as small perturbations, the field-dressed Bloch states ${\tilde u}_{n{\bm k}}$ are given by Eq.~\eqref{field_state}. Using them, the field-corrected interband Berry connections and velocity matrix elements through second order in fields can be written as  
\begin{subequations}
\bea
\tilde{\bm r}_{nm} &=& {\bm r}_{nm} + {\cal E} {\bm r}_{nm}^{\cal E} + {B} {\bm r}_{nm}^{ B} + {\cal E}{B} {\bm r}_{nm}^{{\cal E} B}~,~~ \\
\tilde{\bm v}_{nm} &=& {\bm v}_{nm} + {\cal E} {\bm v}_{nm}^{\cal E} + {B} {\bm v}_{nm}^{ B} + {\cal E}{B} {\bm v}_{nm}^{{\cal E} B}~.~~
\eea
\end{subequations}
Here, the superscript denotes the field-corrected component of the corresponding quantity. These are not bare velocities or interband Berry connections, since they already include field corrections. The $\cal E$-induced velocity matrix element is obtained by evaluating the expectation value of $\hat{\cal H}_0$ with the modified wave function in Eq.~\eqref{field_state}~\cite{Justin_25},
\be \label{v_E}
{\bm v}_{nm}^{\cal E} = \sum_{q\neq n} p_{nq} {\bm v}_{qm}/\varepsilon_{nq} + \sum_{l \neq m} p_{lm} {\bm v}_{nl}/\varepsilon_{ml}~.~~
\ee 
Here, $p_{nq}$ denotes the off-diagonal matrix element of the layer-polarization operator $\hat p$, and $\varepsilon_{nq} = \varepsilon_{n} - \varepsilon_{q}$. Using this result, we obtain
\be \label{r_E}
{\bm r}_{nm}^{\cal E} = -i \hbar \left[  {\bm v}_{nm}^{\cal E} \varepsilon_{n m} + {\bm v}_{nm} 
({p}_{n} - {p}_{m}) \right]/\varepsilon_{nm}^2 ~.~~
\ee 
Similarly, the magnetic field-induced corrections ($ {\bm r}_{nm}^{ B}$ and ${\bm v}_{nm}^{ B}$) are obtained by replacing the interband matrix element of $\hat p$ by $\hat m$ in Eqs.~\eqref{v_E} and \eqref{r_E}, with an overall minus sign. The interband matrix element of $\hat{\bm m}$ is~\cite{Justin_25}
\be
\langle u_{n\bm k}|\hat{\bm m} |u_{p\bm k} \rangle = -(ie\hbar/4) \sum_{l\neq n,p} \left[(1/\varepsilon_{ln} + 1/\varepsilon_{lp}) {\bm v}_{nl} \times {\bm v}_{lp}\right]~.
\ee
Using Eqs.~\eqref {v_E} and \eqref{r_E}, together with their magnetic field-induced counterparts in Eq.~\eqref{gen_BC_OMM}, the first-order corrections to the Berry curvature and orbital magnetic moment can be obtained from the following
\begin{subequations}
\bea
{\bm \Omega}_{n{\bm k}}^{{\cal E}/B} &=& i \sum_{m \neq n} \left( {\bm r}_{nm}^{{\cal E}/B} \times {\bm r}_{mn} + {\bm r}_{nm} \times {\bm r}_{mn}^{{\cal E}/B}\right)~, ~~~~\\
{\bm m}_{n {\bm k}}^{{\cal E}/B} &=& \frac{e}{2}
\sum_{m \neq n} \left[
{\bm v}_{nm}^{{\cal E}/B} \times {\bm r}_{mn} + {\bm v}_{nm} \times {\bm r}_{mn}^{{\cal E}/B} \right]~. ~~~~
\eea
\end{subequations}
These expressions determine the layer polarizability ($\cal E$-induced) and orbital polarizability ($B$-induced) Berry curvatures, as well as the orbital magnetic moments.
The second-order energy correction follows from $\langle {\tilde u}_{n\bm k} | \partial_{\bm k} \hat{\cal H}_0 | {\tilde u}_{n\bm k} \rangle$ by collecting the term proportional to ${\cal E}B$. Specifically, $\varepsilon^{{\cal E}B}_n$ is given by
\be
\varepsilon^{{\cal E}B}_n = - m_{n}^{\cal E}  + 2 \,{\rm Re}  \sum_{l\neq n} \frac{p_{nl} m_{ln}}{\varepsilon_{nl}} ~. ~~
\ee 

To determine the second-order corrections to the Berry curvature and orbital magnetic moments, we need ${\bm v}^{{\cal E}B}_{nm}$ and ${\bm r}^{{\cal E}B}_{nm}$. The second-order correction to the velocity matrix elements follows by collecting the terms proportional to ${\cal E}B$ in $\langle {\tilde u}_{n\bm k} | \partial_{\bm k} \hat{\cal H}_0|{\tilde u}_{m\bm k} \rangle$, giving
\be \label{v_EB}
{\bm v}^{{\cal E}B}_{nm} = - \sum_{q\neq m} \sum_{l\neq n}  \left[ \frac{p_{mq} m_{ln} {\bm v}_{ql}}{\varepsilon_{mq} \varepsilon_{nl}}  + \frac{p_{ln} m_{mq} {\bm v}_{ql}}{\varepsilon_{mq} \varepsilon_{nl}} \right]~. ~~
\ee
Using this result, ${\bm r}^{{\cal E} B}_{nm}$ can be evaluated from ${\bm r}_{nm} = - i \hbar {\bm v}_{nm} / (\varepsilon_n - \varepsilon_m)$ by again collecting the terms proportional to ${\cal E}B$,
\bea \label{r_EB}
&& {\bm r}_{nm}^{{\cal E}B}  = \frac{\hbar}{i} \left[ \frac{{\bm v}_{nm}^{{\cal E}B}}{\varepsilon_{nm}} + \frac{{\bm v}^{\cal E}_{nm} (m_n - m_m)}{\varepsilon^2_{nm}} - \frac{{\bm v}^{B}_{nm} (p_n - p_m)}{\varepsilon^2_{nm}} \right. \nn \\ &&  \left.- \frac{{\bm v}_{nm} (\varepsilon^{{\cal E}B}_n - \varepsilon^{{\cal E}B}_m)}{\varepsilon^2_{nm}}  
 - 2{\bm v}_{nm} \frac{(p_n - p_m) (m_n -m_m)}{\varepsilon_{nm}^3} \right]~.~~~~
\eea
Here, the subscript $m$ denotes the band index. Using Eqs.~\eqref{v_EB} and \eqref{r_EB}, we obtain the second-order correction representing the layer-orbital polarizability of the Berry curvature and orbital magnetic moment,
\begin{subequations}
\bea 
{\bm \Omega}^{{\cal E}B}_n &=& i \sum_{m\neq n} \left[  {\bm r}_{nm}^{{\cal E}B} \times {\bm r}_{mn} + {\bm r}_{nm} \times {\bm r}_{m n}^{{\cal E}B} + {\bm r}^{\cal E}_{nm} \times {\bm r}^{B}_{m n}  \right. \nn\\
&& \left. + {\bm r}^{B}_{nm} \times {\bm r}^{\cal E}_{m n}  \right] ~, ~~ \\
{\bm m}^{{\cal E}B}_n &=& \frac{e}{2} \sum_{m\neq n} \left[ {\bm r}^{{\cal E}B}_{nm} \times {\bm v}_{m n} + {\bm r}_{nm} \times {\bm v}^{{\cal E} B}_{m n} + {\bm r}^{\cal E}_{nm} \times {\bm v}^{B}_{m n}  \right. \nn \\
&& \left. + {\bm v}^{B}_{nm} \times {\bm r}^{\cal E}_{m n}  \right] ~. ~~
\eea 
\end{subequations}
Together, these expressions determine the intrinsic magnetoelectric Hall effect. We emphasize that the perturbative description used above applies only when the gate electric and magnetic field-induced corrections are small compared with the relevant band gap and band-width.

\begin{figure}
    \centering
    \includegraphics[width=\linewidth]{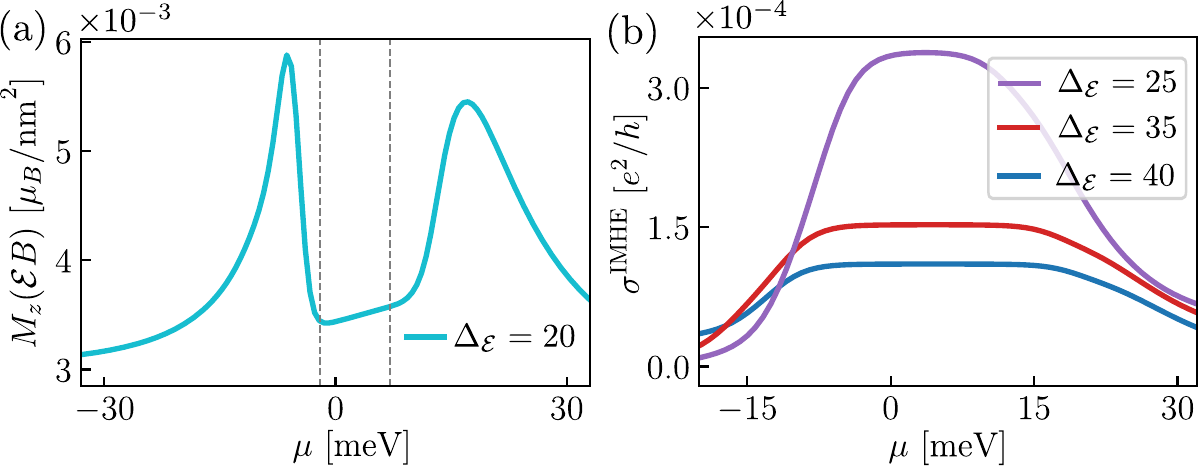}
    \caption{{\bf Orbital magnetization and intrinsic IMHE response.} 
(a) ${\cal E}B$-induced orbital magnetization $M_z({\cal E}B)$ as a function of chemical potential $\mu$ for $\Delta_{\cal E}=20$ meV. The vertical dashed lines indicate the band edges. The magnetization shows pronounced features near the band edges and varies smoothly across the gap. 
(b) Corresponding Fermi-sea component of $\sigma^{\rm IMHE}$ for different gate potential $\Delta_{\cal E}$. The magnitude of the constant in-gap conductivity varies with the applied gate potential.}
    \label{Fig4}
\end{figure}

\textit{Appendix B: Orbital magnetization and magnetoelectric Hall response---}Here, we derive the orbital magnetization underlying the magnetoelectric Hall response proportional to ${\cal E}B$. The intrinsic Hall conductivity can be expressed in terms of the orbital magnetization as~\cite{jackson_12, Wen_prr25}
\begin{equation}
\sigma_{ab} = j_a/E_b = -e\, \epsilon_{abc} \, (\partial_{\mu} M_c)~,
\label{M_to_j}
\end{equation}
where $\mu$ is the chemical potential. To obtain a current $j \propto E {\cal E} B$, we therefore require an orbital magnetization $M_z \propto {\cal E}B$.

The orbital magnetization follows from the free energy via $M = -\partial F/\partial B$. We therefore evaluate the free energy up to order ${\cal E}B^2$. The free energy is given by~\cite{xiao_rmp10_berry}
\begin{equation}
F = - \frac{1}{\beta} \int_{n\bm k} {\cal D}^{-1} \,
\ln \left[1 + e^{-\beta(\tilde{\varepsilon} - \mu)} \right]~,
\label{free_enrg_gen}
\end{equation}
where ${\cal D} = \left[1+(e/\hbar) {\bm \Omega} \cdot {\bm B} \right]^{-1}$ is the phase-space correction factor~\cite{Xiao_prl05}. Defining $
g(\tilde{\varepsilon}) = -\dfrac{1}{\beta} \ln \left[ 1 + e^{-\beta (\tilde{\varepsilon} - \mu)} \right]$,
we can expand $g(\tilde{\varepsilon})$ to the required order
\bea
g(\tilde{\varepsilon}) &\simeq&  g(\varepsilon)
+ \left(B\varepsilon^{B} + {\cal E}\varepsilon^{\cal E} + {\cal E}B \varepsilon^{{\cal E}B}\right)\partial_\varepsilon g(\varepsilon) \nonumber \\
&& + {\cal E}B\, \varepsilon^{\cal E}\varepsilon^{B}\, \partial^2_\varepsilon g(\varepsilon)~.
\eea
The free energy, including field-induced corrections to the phase-space measure, is
\begin{equation}
F = \int_{n\bm k}
\left[ 1 + \frac{e}{\hbar}
\left( \Omega + {\cal E} \Omega^{\cal E} + B \Omega^{B}
+ {\cal E}B \Omega^{{\cal E}B} \right)  B
\right] g(\tilde{\varepsilon})~.
\end{equation}
Collecting terms proportional to ${\cal E}B^2$, we obtain the orbital magnetization
\bea
M_z({\cal E}B) &=& -\frac{e}{\hbar} {\cal E} B \int_{n \bm k} \Big[
\Omega^{{\cal E}B} g(\varepsilon)
+ \left(\Omega\, \varepsilon^{{\cal E}B}
+ \Omega^{\cal E} \varepsilon^B \right. \nonumber \\
&& \left. + \Omega^{B} \varepsilon^{\cal E} \right)\partial_\varepsilon g(\varepsilon)  + \Omega \, \varepsilon^{\cal E} \varepsilon^B \, \partial^2_\varepsilon g(\varepsilon)
\Big]~.
\label{Mz_final}
\eea
The first two terms in the above equation [also presented in Eq.~\eqref{Orb_Mag_EB}] arise from the coupled ${\cal E}B$ correction to the geometric quantities.
Using $\partial_\mu g(\varepsilon) = -f_0(\varepsilon)$ and $\partial_\mu f_0 = -\partial_\varepsilon f_0$, we obtain the intrinsic Hall current from Eq.~\eqref{M_to_j} as
\bea \label{j_full}
j_a &=& - \frac{e^2}{\hbar} \epsilon_{abz} E_b {\cal E} B \int_{n\bm k}
\Big[
\Omega^{{\cal E}B} f_0
+ \Omega \varepsilon^{{\cal E}B} \, \partial_{\varepsilon}f_0 \nonumber \\
&&+ \left(\Omega^B \varepsilon^{\cal E}
+ \Omega^{\cal E} \varepsilon^B \right)\partial_\varepsilon f_0 + \Omega \varepsilon^{\cal E} \varepsilon^B \, \partial^2_\varepsilon f_0
\Big]~.~~~
\eea
Equation~\eqref{j_full} gives the complete intrinsic magnetoelectric Hall current, including all contributions up to order ${\cal E}B$. This expression is fully consistent with that obtained independently from the semiclassical Boltzmann formalism (see Sec.~S1 of SM~\cite{Note1}), establishing the equivalence between the orbital magnetization and semiclassical transport approaches. More generally, it shows that, as in the anomalous Hall effect~\cite{xiao_rmp10_berry, Wen_prr25}, field-induced Fermi-sea responses can be interpreted in terms of orbital magnetization.

We present the ${\cal E}B$-induced orbital magnetization $M_z({\cal E}B)$ as a function of $\mu$ in Fig.~\ref{Fig4}(a), for ${\cal E} \approx 3.5 \times 10^6$ V/m and $B = 0.1$ T. The magnetization shows pronounced variations near the band edges and varies monotonically across the gap. The magnitude of $M_z({\cal E}B)$ reaches values of order $\sim 0.005\,\mu_B/{\rm nm}^2$, which should be detectable using optical Kerr-effect measurements. Figure~\ref{Fig4}(b) shows the Fermi-sea component of $\sigma^{\rm IMHE}$, which remains finite and nonquantized in the band-gap region. This nearly constant in-gap response is consistent with the monotonic variation of $M_z({\cal E}B)$ in the gap, since $\sigma^{\rm IMHE} \propto \partial_\mu M_z({\cal E}B)$.

Finally, we emphasize that this mechanism does not rely on symmetry-protected topological edge states and therefore yields a non-quantized Hall response. Its magnitude can be tuned by material parameters such as gate voltage and disorder, as shown in Fig.~\ref{Fig4}(b); nonetheless, its geometric origin remains robust in sufficiently clean samples.

\end{document}